\documentstyle[aps,preprint]{revtex}
\begin{document}
\draft
\title{Temperature Dependence of the  Optical Response of
       Small Sodium Clusters}

\author{Aurel Bulgac and Caio Lewenkopf}

\address{Department of Physics, FM--15, University of Washington,
         Seattle, WA 98195, USA\\}
\maketitle
%
\begin{abstract}     
We present an analysis of the temperature dependence of the optical
response of small sodium clusters in a temperature range bracketing the
melting phase transition. When the temperature increases, the mean
excitation energy undergoes a red shift and the plasmon is significantly
broadened, in agreement with recent experimental data. We show that the
single--particle levels acquire a prominent width and the HOMO--LUMO
gap as well as the width of the occupied band are reduced due to large
thermal cluster size and shape fluctuations.
This results in a sharp increase of the static polarizability with
temperature.
%
%
\end{abstract}
\pacs{PACS number:  36.40.+d, 73.20.Mf, 31.20.Pv, 36.20.Kd}
%

\narrowtext

One of the most intriguing questions in atomic cluster physics is the
understanding of the relative importance of electronic versus ionic
degrees of freedom for different observables. There is clear
experimental evidence for the formation of either geometric or
electronic shells for sodium clusters with $N\approx 1500 \dots  2000$
atoms \cite{TPM90} in the abundance spectra, depending on the cluster
temperature.  At the present time there is only a qualitative
theoretical understanding of this experimental result \cite{SB92}. Here
the temperature influence on the ionic degrees of freedom, which
dominate the  thermodynamics, was not yet investigated and thus the
question is still not satisfactorily understood. By measuring the
optical response of small alkali clusters as a function of temperature
the Freiburg group \cite{Hab93} provided further input from a slightly
different perspective: The photo--depletion spectra show a strong
temperature dependence of the plasmon peak and its broadening. In this
work we show that these spectra can be understood quantitatively in a
framework that includes electrons and ions explicitly. In qualitative
terms the scenario is very intuitive: with increasing temperature the
system becomes less rigid and expands. A naive estimate based on bulk
values would discard this explanation since it leads to a relatively
small effect. Our calculations, however, show surprisingly large
expansion coefficients and fluctuations.

We discard the two customary approaches for the description of the
electronic spectrum of alkali clusters as suitable for a discussion of
temperature effects: The jellium model, which is very
successful in describing the electronic shell effects and which leads to
a qualitatively accurate picture of the electronic excitations
\cite{deH93}, does not incorporate ions explicitly.
Any shape and size dynamics has to be included by hand and driven by
the bulk parameters.
Much effort has also been devoted to a different approach: very
sophisticated  {\sl ab initio} calculations for $N \leq 20$ atoms
\cite{KFK91}.
This approach can provide rich information about ground state (and few
isomeric) ionic configurations.
The question is: How relevant is this information for understanding
experimental data taken around the bulk melting temperature (for Na,
$T_m=371$ K)?

A partial answer to this question was given in Ref. \cite{BJ}, where the
ions were treated explicitly and the electrons only in a implicit manner.
With increasing temperature, while the cluster melts, its geometrical
properties change dramatically.
A cluster undergoes a significantly more pronounced thermal expansion
than the bulk, and its thermal shape and size fluctuations are very
large.
Since the gross features of the optical response of a metallic cluster
(the line shape and position) are dominated by its geometry, it was
suggested in Ref. \cite{BJ} that the melting can be indirectly put in
evidence by studying the temperature dependence of the Mie resonance.
With the exception of Refs. \cite{WLB93}, all calculations of the electronic
excitations were performed at zero temperature, which is definitely not
the experimental situation.
A model aiming to address temperature effects requires an explicit
treatment of atoms and electrons.
However, any {\sl ab initio} scheme encompassing the coupling to the
environment and the electronic excitations is a quite daunting task
\cite{and}.
This invites for a somewhat simplified approach.

The approach we have chosen is based on a distance dependent
H\"uckel/tight--binding  parametrization of the LDA Hamiltonian
\cite{PS92}, which includes effectively the $3s$ and $3p$ atomic
orbitals of an isolated sodium atom. The electronic Hamiltonian reads
\begin{eqnarray}
h_{ii}  &=&  \sum_{k \neq i} \rho (R_{ik}) \nonumber \\
h_{ij}  &=&  t_{ss}(R_{ij}) - \sum_{k \neq i,j}
           \left[ \frac{t_{sp}(R_{ik})\, t_{sp}(R_{jk})}
                       {e_{3p} - e_{3s}}\,
                  \frac{{\bf R}_{ik}\cdot {\bf R}_{jk}}{R_{ik}R_{jk}}
           \right]
\end{eqnarray}
where $R_{ij}$ is the distance between the ions $i$ and $j$ and $\rho$,
$t_{sp}$ and $t_{ss}$ are distance dependent fitted functions.
The electronic wave functions
\begin{equation}
  \phi_\mu({\bf r}) = \sum _{k} C_\mu^k \,\varphi _{3s} ({\bf r-R}_k)
\end{equation}
are linear combinations of $3s$ atomic orbitals $\varphi_{3s}({\bf r-R}_k)$
centered at sites ${\bf R}_k$, where the coefficients $C_\mu^k$ are determined
by the eigenvalue equation $\sum_l h_{kl}C^l_\mu = \varepsilon_\mu C^k_\mu$.
The parameterization Ref. \cite{PS92} leads to ground state ionic
configurations in remarkable good agreement with quantum chemistry
calculations \cite{KFK91}.
The electrons are considered always in their lowest Born--Oppenheimer
state and the interionic forces are given by the Hellman--Feynman force.
The coupling of the ionic degrees of freedom to a thermal bath is
simulated as in Ref. \cite{BJ}.
We discuss temperatures which are either comparable or higher than the
Debye temperature for bulk sodium ($T_D=150$ K), where one can safely
use a classical description for the ionic degrees of freedom.
Presently the experiments are done at $T>100$ K.
The relative simplicity of this electronic Hamiltonian allows us to
perform rather long canonical molecular dynamics simulations of small
sodium clusters (at least $10^{-9}$ s for each temperature).
Long trajectories are desirable, since the structural changes occuring
in a cluster while it undergoes a phase transition are relatively slow
in time.
This is the main advantage of the present approach over an {\sl ab initio}
type of calculation, where a reliable description of the phase transition
in a cluster is nowadays computationally prohibitive.
Along each molecular dynamics trajectory, for each different spatial
configuration of a cluster, we have performed RPA calculations of the
optical response.
In this way, at any finite temperature the cluster properties we are
presenting are proper canonical averages.
The RPA is implemented in the following way \cite{BB91}: The polarization
propagator
\begin{equation}
      \Pi_{RPA} = \Pi_0 + \Pi_0 \, V \, \Pi_{RPA} \; ,
\end{equation}
is given in terms of the residual electron-electron Coulomb interaction
$V(r)=e^2/r$ and the free particle--hole propagator $\Pi_0$
\begin{eqnarray}
  \Pi_{0}({\bf r}, {\bf r'}, \omega) = \sum _{p,h}
     \phi_p({\bf r})\phi_h^*({\bf r})\,
      \frac{2\,(\varepsilon_p -\varepsilon _h)}
           {(\omega +i0^+)^2-(\varepsilon _p-\varepsilon _h)^2}\,
     \phi_p^*({\bf r'})\phi _h({\bf r'})\; .
\end{eqnarray}
The response of the cluster to a weak external field $F({\bf r})$
is given by $S = \mbox{Im}\langle F | \Pi_{RPA} | F \rangle/\pi$.

We have performed calculations for a variety of clusters in the range
$4 \le N \le 21$.
The structural properties of neutral clusters are in good agreement
with the previous analysis \cite{BJ} and partially at odds with some
recent results \cite{Spi94}.
Since this H\"uckel parameterization was optimized for neutral clusters
it should be taken cautiously when addressing charged ones.
When evaporation sets in at $T \approx 500 \cdots 600$ K our method
has a further deficiency since evaporative ensembles are rather
poorly described in canonical molecular dynamics.
Even with these provisos, our results still reproduce nicely the main
features of the experimental data \cite{Hab93} for single positively
charged clusters.

The results for the single--particle spectra can be summarized as
follows: With increasing temperature, the single particle energy levels
acquire a width and the width of the occupied band decreases, see Fig.
1.  The decrease in the band width is mainly due to the significant
thermal expansion of the cluster.  The width of the single particle
energy levels has two origins: {\sl i}) the increase with temperature of
the oscillation amplitude of the ions around their equilibria and the
coupling of the ionic and electronic degrees of freedom (the analogous
of the bulk electron--phonon coupling);
{\sl ii}) the large thermal size and
shape fluctuations of the ionic background.  This last aspect is
characteristic of small metal particles only and is not present in the
bulk. Only the increase in the ionic oscillation amplitude with
temperature, has been so far considered within a schematic random matrix
approach \cite{AK93}. In the insert of Fig. 1 the temperature dependence
of the HOMO-LUMO gap  $\Delta \varepsilon $ distribution $P(\Delta
\varepsilon )$ is shown.  We observe that for $N \le 20$ at temperatures
$T<500$ K, the condition $T\ll \Delta\varepsilon$ is always fulfilled.
This insures the quality of the Born--Oppenheimer approximation.
Though, the appearance of relatively small $\Delta \varepsilon $
certainly raises serious doubts concerning the applicability of this
approximation  to larger clusters.

It is interesting to establish the effect of the explicit treatment of
the electronic degrees of freedom on the properties of a cluster, as its
temperature changes. As expected, the role of electrons is especially
important in magic clusters. In particular Na$_{20}$ is spherical at low
temperatures, while in Ref. \cite{BJ} it was deformed.
Our simulations indicate that with increasing temperature (particularly
above melting) the stabilizing role of the electronic shells is very
much reduced: the studied clusters become deformed and the shape and
size fluctuations are significant.
This is a rather disquieting observation, since the current understanding
of the electronic shell effects assumes a melted ionic background.

Experimental data \cite{Hab93} indicate that the optical response of
sodium clusters is significantly modified as a function of temperature.
At low temperatures (around 100 K)  most of the oscillator strength is
concentrated in a few relatively close and sharp lines, in the vicinity
of the predicted Mie resonance.
At higher temperatures the mean excitation energy of the response moves
towards smaller excitation energies and different lines merge into a wide
structure.
Both of these trends are reproduced in our calculations, see Figs. 2 and 3.

There are several possible mechanisms that can lead to the broadening
of the plasmon.
One can almost immediately dismiss the coupling to two particle--two
hole states or to the continuum as important mechanisms, since
the density of electronic $2p$--$2h$ configurations and the escape
width for $N \le 20$ is insignificant in the region of the Mie
resonance.
The Landau fragmentation, on the other hand, certainly plays some
role (and seems to be more pronounced in neutral than in charged
clusters) \cite{YBB89}.
The tight--binding model used here does not describe accurately the
unperturbed single--particle excitations in the region of the Mie
resonance and Landau damping plays a minor r\^ole here.
Had we focussed the discussion only on the ionic vibrations around
some equilibrium geometry corresponding to very low temperatures,
our modelling would be unsatisfactory.
Such plasmon broadening mechanism due to electron--phonon coupling is
likely to play a significant role only for very small clusters
\cite{WLB93,BT89}.
Usually the plasmon line shape is linked with the shape of a
cluster \cite{deH93,Ped93}: A single hump is indicative of a spherical
shape, a double hump is related to an axially deformed cluster and
three humps are associated with a triaxial shape \cite{Ped93}.
The presence of several peaks cannot be uniquely linked with the cluster
deformation, since departures from the simple jellium picture and the
presence of shape isomers can have a similar effect.
In addition there are also notable differences between the optical
response of a neutral and charged clusters with the same number of
electrons, see Fig. 2 and Refs. \cite{Hab93,PWK91}.
The measured response of a hot Na$_{20}$, see Ref. \cite{PWK91}, has
a double hump line shape, similar to Fig. 3, while the measured line
shape for Na$_{21}^{+}$ exhibits a single hump \cite{Hab93}.

We advocate that the main mechanism responsible for the measured line
width \cite{Hab93} is the following:
Above melting temperature the ions become extremely mobile \cite{BJ}
and the cluster is characterized by significant shape fluctuations
and not only by a static deformation.
The photoexcitation is a rapid process at the ionic time scale and the
response of a cluster will reflect the instantaneous geometrical arrangement
of the ions.
The measured spectrum is thus a properly weighted response of all
allowed shapes.
Therefore, the observed line shape is not an intrinsic width of the
electronic excitation, but an ensemble averaged image of the thermal
shape fluctuations.
Our calculations are for neutral clusters, while experiment \cite{Hab93}
was performed on positively ionized clusters.
Even though the present formalism is not well suited for charged clusters,
our results suggest that both at low and high temperatures one can account
completely for the line width by considering only the electron--phonon
coupling and the thermodynamics of the cluster, all other effects playing
a minor role only.

Even though we do not describe the details of the experimental results
(exact position of peaks and overall line shape), the main temperature
effects are well reproduced: the magnitude of the red shift of the
average excitation energy and the magnitude of the increase of the line
width with increasing temperature.
This is confirmed also by the strong temperature dependence of
the polarizability of these clusters, which can be independently
measured.
The broadening of the single--particle levels could in principle be
seen in photoionization spectra.
The size of the temperature effects we report on is significantly
larger than one would have expected  from the study of similar bulk
properties.
This shows once again that atomic clusters are in more respects rather
unique objects.

We express our thanks for useful discussions to G.F. Bertsch, S.
Bj{\o}rnholm and D. Tom\'anek and in particular to N. Ju.
We also thank F. Spiegelmann for sending us additional tight--binding
parameters \cite{PS92}.
Financial support from the National Science Foundation and the
Department of Energy is greatly appreciated.


\vskip3.5truecm
\noindent
{\bf Figure Captions}
\vskip1.0truecm
\noindent
FIG.1 Probability distribution $P(\varepsilon_{s.p.})$ of single-particle
      energies in Na$_{20}$ for $T=150,350$ and 450 K.
      Insert: gap energy distribution $P(\Delta\varepsilon)$ for the same
      temperatures. An increase in $T$ corresponds to broader distribution.

\vskip0.5truecm
\noindent
FIG.2 Oscillator strength for Na$_8$ as a function of energy $\omega$
      for $T=150, 350$ and 450 K.
      Insert: the same for Na$_9^+$, $T=150$ and 450 K. An increase in
      $T$ causes a red shift in the response.

\vskip0.5truecm
\noindent
FIG.3 Oscillator strength for Na$_{20}$ as a function of energy $\omega$
      for $T=200, 300$ and 400 K.
      Insert: polarizability $\alpha$ (normalized to one atom $\alpha_1$
      \cite{deH93}) as a function of the r.m.s. radius $r$ for the same
      values of $T$.
      Increasing values of $r$ correspond to increasing $T$. Error bars
      correspond to the variance of the respective quantities due to
      thermal fluctuations.

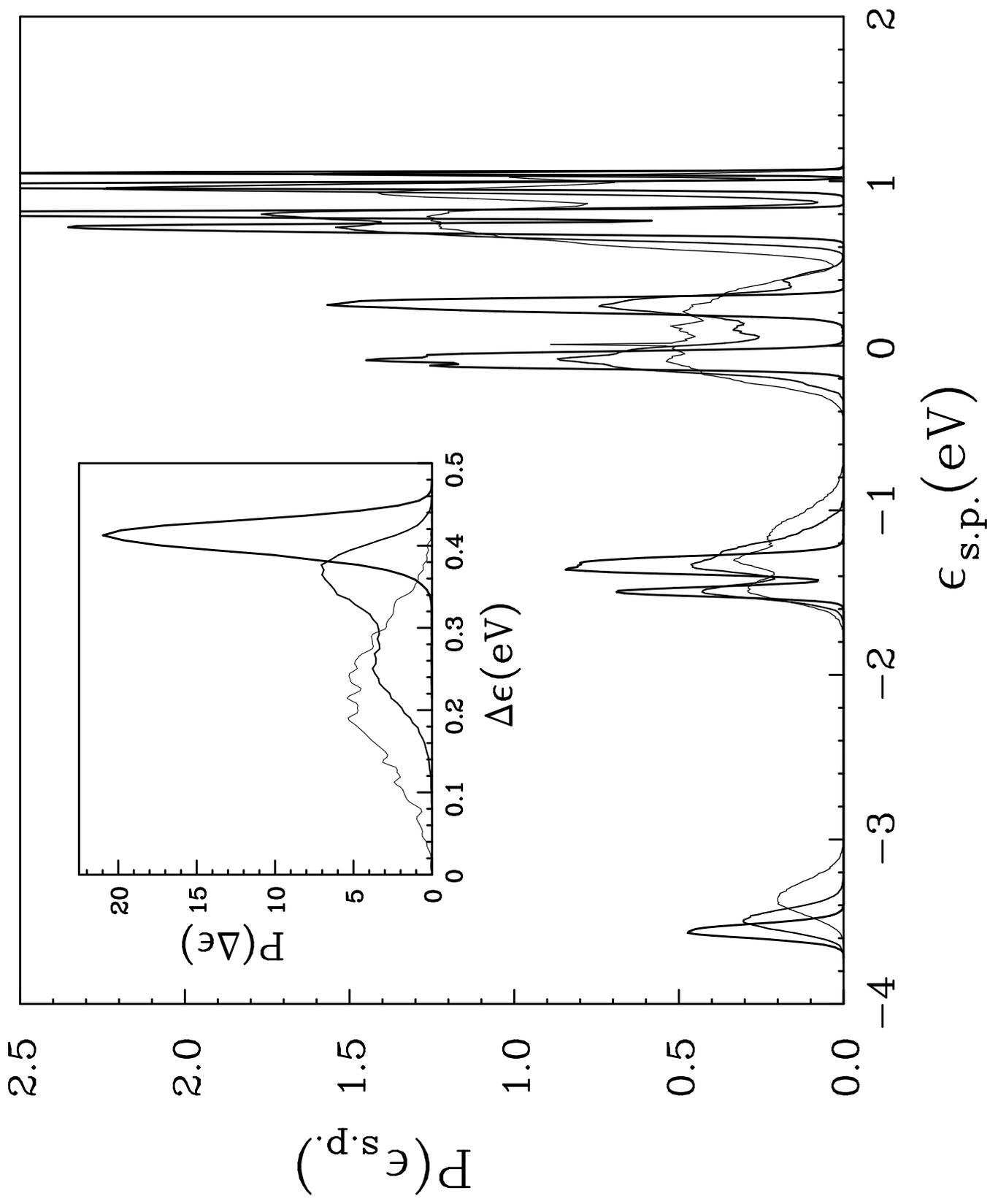
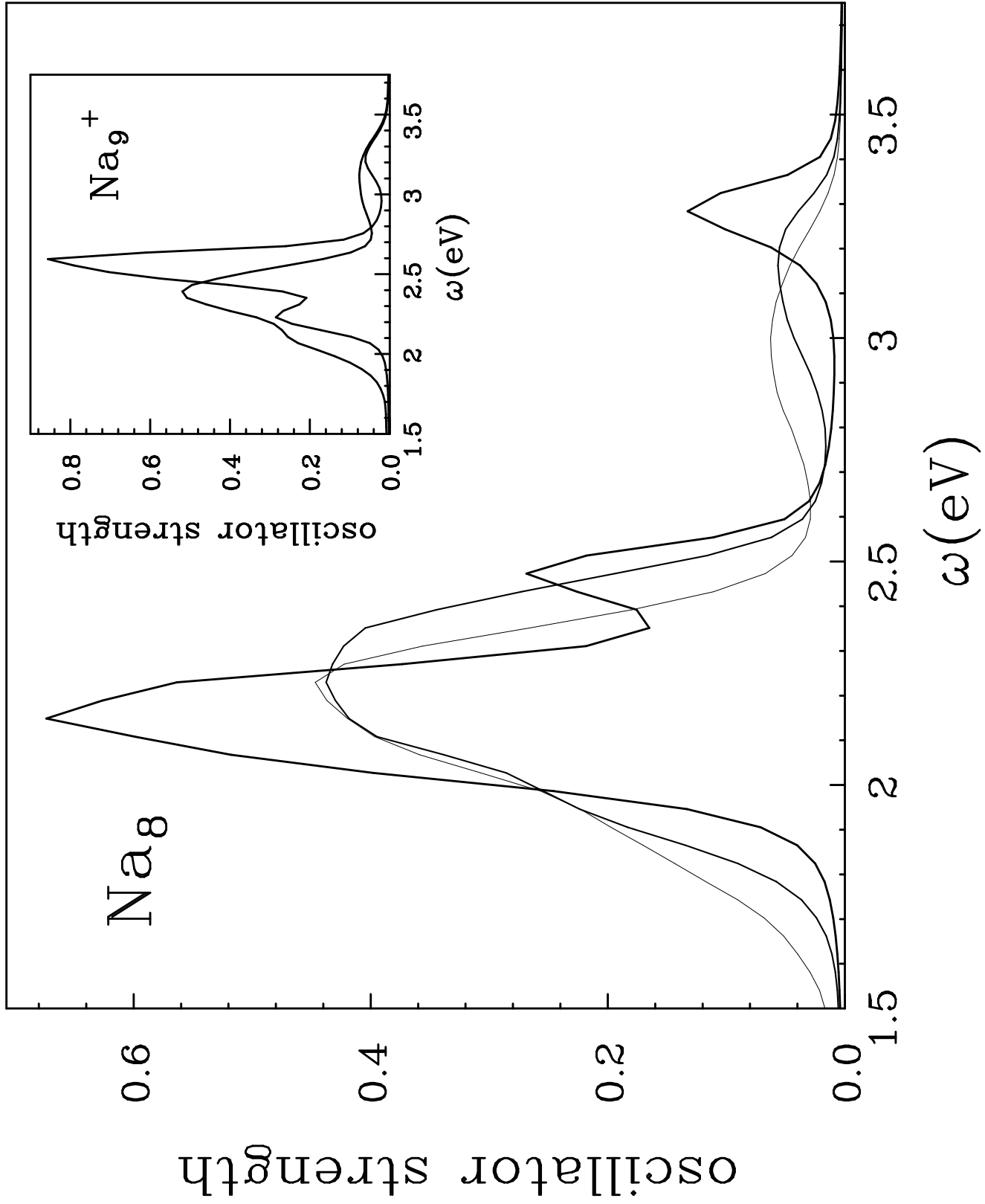
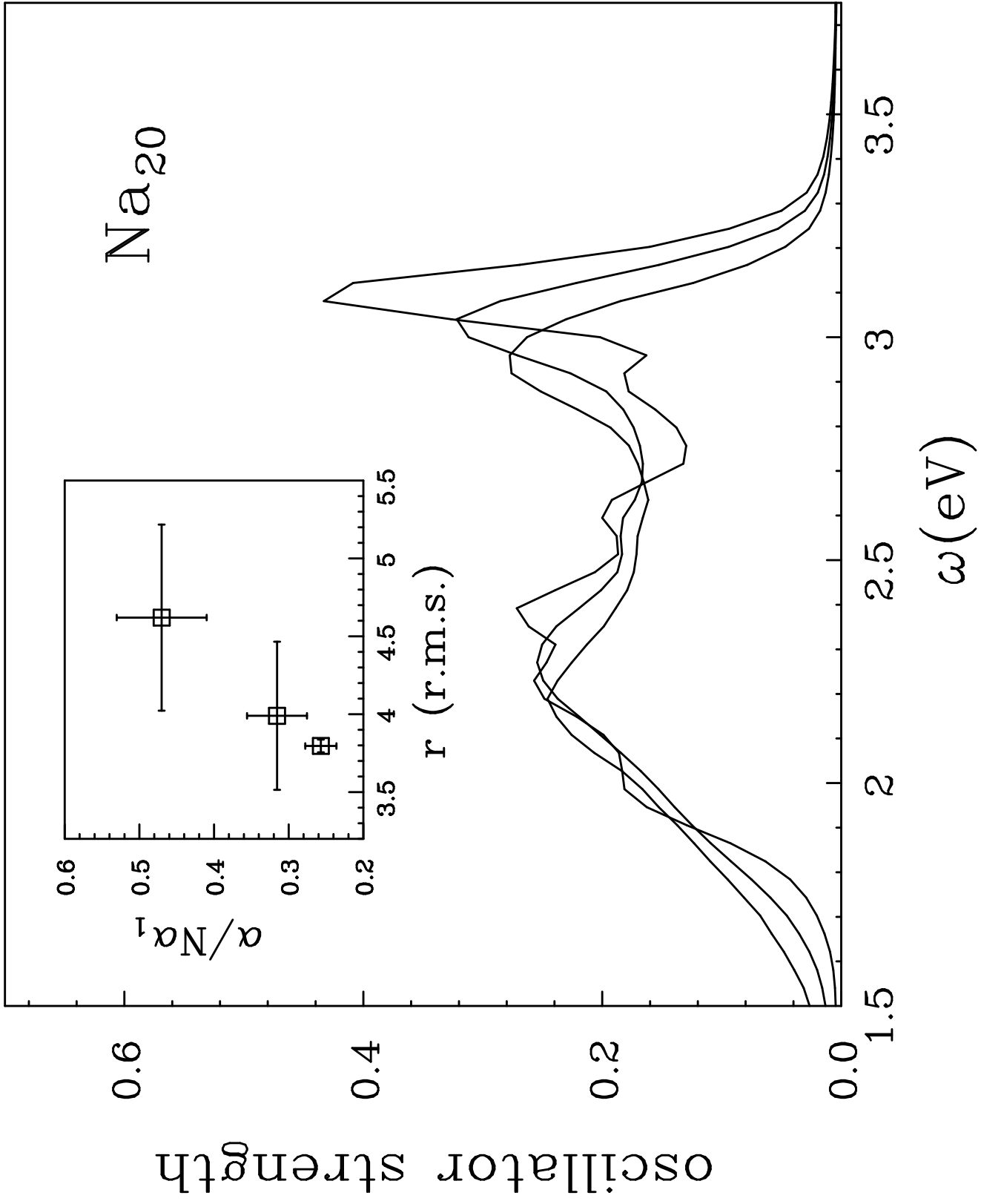
\end{document}